\documentclass[showpacs,preprintnumbers,amssymb]{revtex4-2}
\AtBeginDocument{
  \setlength\abovedisplayskip{0pt}
  \setlength\belowdisplayskip{0pt}}
\usepackage{graphicx}
\usepackage{dcolumn}
\usepackage[dvipsnames]{xcolor}
\usepackage{bm}
\usepackage{amsmath}
\usepackage{amssymb}
\usepackage{epsfig}
\usepackage{amsfonts}
\usepackage{lineno,hyperref}
\usepackage{array}
\usepackage{float}
\usepackage{microtype}
\usepackage{multirow}
\usepackage{adjustbox}
\usepackage[english]{babel}
\usepackage{epstopdf}
\usepackage{blindtext}
\usepackage{subcaption}
\usepackage[a4paper, total={6.5in, 10in}]{geometry}

\def \a{\alpha}
\def \b{\beta}
\def \l{\lambda}

\def \e{\epsilon}

\def \k{\kappa}

\def \be{\begin{equation}}
\def \ee{\end{equation}}
\def \ben{\begin{eqnarray}}
\def \een{\end{eqnarray}}
\def \o{\omega}

\def \p{\partial}

\def \M{\mathcal{M}}

\def \G{\bar{G}}

\def \k{\kappa}
\def \R{\bar{R}}
\def \T{\bar{T}}

\def \La{\mathcal{L}}
\def \T{\bar{T}}

\DeclareUnicodeCharacter{2212}{-}

\begin{document}

\title{Collapsing scenarios of K-essence generalized Vaidya spacetime under $f(\bar{R},\bar{T})$ gravity}    

\author{Arijit Panda}
\email{arijitpanda260195@gmail.com}
\affiliation{Department of Physics, Raiganj University, Raiganj, Uttar Dinajpur, West Bengal, India, 733 134. $\&$\\
Department of Physics, Prabhat Kumar College, Contai, Purba Medinipur, India, 721 404.}

\author{Goutam Manna$^a$}
\email{goutammanna.pkc@gmail.com}
\altaffiliation{$^a$Corresponding author}
\affiliation{Department of Physics, Prabhat Kumar College, Contai, Purba Medinipur 721404, India $\&$\\ Institute of Astronomy Space and Earth Science, Kolkata 700054, India}

\author{Saibal Ray}
\email{saibal.ray@gla.ac.in}
\affiliation{{Centre for Cosmology, Astrophysics and Space Science (CCASS), GLA University, Mathura 281406, Uttar Pradesh, India}}

\author{Maxim Khlopov}
\email{khlopov@apc.in2p3.fr}
\affiliation{Institute of Physics, Southern Federal University, 194 Stachki, Rostov-on-Don 344090, Russian Federation $\&$\\
National Research Nuclear University, MEPHI, Moscow, Russian Federation $\&$\\ Virtual Institute of Astroparticle Physics 10, rue Garreau, 75018 Paris, France}

\author{Md. Rabiul Islam}
\email{rabi76.mri@gmail.com}
\affiliation{Department of Physics, Raiganj University, Raiganj, Uttar Dinajpur-733134, West Bengal, India.}

\begin{abstract} 
The paper investigates the collapse of the generalized emergent Vaidya spacetime in the setting of $f(\bar{R}, \bar{T})$ gravity, specifically in K-essence theory. In this study, the Dirac-Born-Infeld type non-standard Lagrangian is used to calculate the emergent metric $\bar{G}_{\mu\nu}$, which is not conformally equivalent to the conventional gravitational metric. We use the function $f(\bar{R}, \bar{T})$ to reflect the additive nature of the emergent Ricci scalar ($\bar{R}$) and the trace of the emergent energy-momentum tensor ($\bar{T}$). Our study demonstrates that certain choices of $f(\bar{R}, \bar{T})$ may result in the existence of a naked singularity caused by gravitational collapse. The alternative $f(\bar{R}, \bar{T})$ values resulted in an accelerating universe dominated by dark energy. Moreover, the investigation showed the presence of both positive and negative masses, which might suggest the coexistence of dark matter and dark energy. Furthermore, for a given quantity of kinetic part of the K-essence scalar field, mass is completely changed into energy, meaning that spacetime is Minkowskian. The K-essence theory may also be employed as a dark energy framework and a basic gravitational theory, making it possible for researchers to investigate a wide ranges of cosmic phenomena.
\end{abstract}


\keywords{Gravitational collapse, Naked singularity, Dark energy, K-essence, Modified theories of gravity, Vaidya geometry}

\pacs{04.20.-q, 04.20.Dw, 04.50.Kd,  04.70.Bw}  

\maketitle

\section{Introduction}

Gravitational collapse is a well-recognized phenomenon in the fields of general relativity and astrophysics, as evidenced by the works of Joshi et al.~\cite{Joshi1,Joshi2}. It serves as a crucial avenue for comprehending many astrophysical facets inside our universe. The phenomenon of gravitational collapse offers valuable insights into several aspects of astrophysics, such as structure development, stellar characteristics, black hole generation, and the formation of white dwarfs or neutron stars, among other phenomena. Specifically, gravitational collapse is an event where a star undergoes the process of collapse due to its mass, and depending upon some specific condition of the initial mass it may end up at different stages of its collapse. Sometimes, the process of stellar collapse bypasses the intermediate phases, such as the production of a white dwarf or neutron star, and proceeds directly to the creation of a singularity known as a black hole (BH), provided that the star possesses a substantial mass exceeding $20M_{\odot}$, where $M_{\odot}$ denotes the solar mass. In their investigation of gravitational collapse, Oppenheimer and Snyder employed a spherically symmetric dust cloud as the fundamental framework. The process of collapse, as depicted in their study, indicated the existence of a black hole in the final state. This model was characterised by a static Schwarzschild exterior and Friedmann interior, serving as an idealised representation of the phenomenon \cite{Oppenheimer}. Subsequently, extensive research pertaining to gravitational collapse has been conducted globally, and comprehensive evaluations of these studies can be found in the references \cite{Joshi1,Joshi2,Malafarina}.

In the context of this framework, Roger Penrose proposed the renowned Cosmic Censorship Hypothesis (CCH), which argues that a cosmic singularity will invariably be surrounded by an event horizon \cite{Penrose1,Penrose2}. Upon further examination of this hypothesis, a novel model has emerged in which the singularity is not concealed by an event horizon, resulting in the formation of a ``naked singularity'' \cite{Eardley,Christodoulou,Newman,Dwivedi,Joshi3,Joshi4,Joshi5,Waugh,Ori,Lake,Szekeres}. This concept is considered crucial in the advancement of a viable quantum gravity theory. The first relativistic line element, which accurately depicted the spacetime of a plausible star, was introduced by P. C. Vaidya in 1951 \cite{Vaidya}. It generalized the static solution of Schwarzschild by representing the radiation for a non-static mass. The solution proposed by Schwarzschild represents the spacetime surrounding a spherically symmetric, cold, black object with a fixed mass. Hence, it is evident that the model is unable to adequately represent spacetime beyond the boundaries of a star. The solution proposed by Vaidya \cite{Vaidya}, commonly referred to as Vaidya spacetime or the radiating Schwarzschild metric, was presented as a potential explanatory framework. It is important to acknowledge that the primary difference between the two metrics lies in the fact that the Vaidya metric incorporates a time-dependent mass parameter, whilst the Schwarzschild metric utilizes a constant mass value, hence resulting in a spacetime that varies with time. 

Husain \cite{Husain} describes the generalized Vaidya spacetime, which corresponds to the gravitational collapse of a null fluid. Wang and Wu \cite{Wang} expanded this generalization by demonstrating that the energy-momentum tensor for Type-I and Type-II matter fields is proportional to the mass function. Furthermore, generalized Vaidya spacetime features an off-diagonal term that may result in negative energy for a particle \cite{Vertogradov}. For a given mass function, the Generalized Vaidya metric produces a homothetic Killing vector. This extra symmetry may help construct a constant of motion related to both angular momentum and energy \cite{Vertogradov5}.  As a result, the combination of particular solutions by linear superposition fulfills the field equation. In this generalization, the mass function depends on both space ($r$) and time ($v$). Using this generalized Vaidya spacetime, we can investigate many dynamical behavior of the cosmos. The Vaidya metric has been extensively studied in the literatures, and several notable works have contributed to our understanding of this topic. Relevant works on the Vaidya metric can be found in \cite{Rudra1,Heydarzade1,Rudra2,Rudra3,Heydarzade2,Vertogradov1,Vertogradov2,Vertogradov3,Vertogradov4}.

On the other hand, the findings derived from type IA Supernovae, Baryon Acoustic Oscillator (BAO), Cosmic Microwave Background (WMAP7), and Planck discoveries have provided compelling evidence indicating that the expansion of our universe is now in the late-time accelerated phase \cite{Riess,Perlmutter,Komatsu,Planck1,Planck2,Planck3}. The cause of this acceleration has been attributed to dark energy, prompting scientists to contemplate modifying the general theory of relativity by incorporating a curvature term as a variable function of $R$ in the action ($R$ referred to as the Ricci scalar). This culminated in the development of the modified theory of gravity \cite{Carroll}. The standard theory of gravity has been modified by various relevant theories, such as $f(R)$, $f(G)$, $f(R, \La_m)$, and $f(R, G)$, where $\La_m$ represents the matter Lagrangian and $G$ denotes the Gauss-Bonnet invariant term \cite{Oikonomou1,Oikonomou2,Sotiriou,Felice,Nojiri1,Nojiri2,Nojiri3,Capozziello,Nojiri4,Elizalde,Cognola,Myrzakulov,Durrer,Copeland,Zubair5,Maurya2022,Maurya2023}. 

As previously mentioned, $f(R)$ gravity is a theoretical framework in which the action is formulated as a function of the Ricci scalar, denoted as $R$. One notable aspect of this updated theory is its ability to account for the acceleration of the cosmos without requiring the addition of a cosmological constant ($\Lambda$). Previous studies in this subject have introduced some generalizations by considering the connection between the Ricci Scalar ($R$) and matter ($\La_m$), as well as the trace of the energy-momentum tensor ($T$), among other factors. The $f(R, \La_m)$ theory was developed by Harko et al. \cite{Harko1}, whereby the authors investigate the coupling between the Ricci scalar ($R$) and the matter Lagrangian ($\La_m$). The theory provides the gravitational field equation and the equation of motion for test particles. Subsequently, Harko et al. \cite{Harko2} extended the $f(R)$ theory by proposing a connection between the Ricci scalar and the trace of the energy-momentum tensor, resulting in the formulation of the $f(R,T)$ theory. In their study, the authors of \cite{Harko2} employed the widely recognized variational technique to get the field equation of this particular theory. Additionally, they performed calculations to determine the covariant divergence of the energy-momentum tensor. In recent years, there has been a rise in the popularity of this idea, leading to a significant body of research and ongoing scholarly endeavours in this field. The study conducted by Sahoo et al. \cite{Sahoo1} investigates the thermodynamics of Bianchi-III and Bianchi-$VI_0$ cosmological models with a string fluid source in the context of $f(R, T)$ gravity. The authors also consider the limitations in the choice of $T$ as discussed in Alvarenga's work \cite{Alvarenga} and the thermodynamics in this theory as examined by Sharif et al. \cite{Sharif1}. Numerous studies have been conducted in the present context \cite{Singh1,Singh2,Baffou,Sahu,Das2017,Deb2018a,Deb2018b,Deb2019a,Deb2019b,Panda2024a,Panda2024b,Sharif2,Houndjo1,Houndjo2,Jamil,Dombriz,Nojiri5,Houndjo3,Moraes,Zubair1,Singh3,Singh4,Clifton,Fradkin,Vasiliev,Khoury1,Khoury2,Vainshtein,Horndeski,Deffayet,Tiago,Rosa,Rej,Sharif3,Jasim}.

Usually, researchers prefer to work with the canonical Lagrangian ($\La=T-V$) to formulate the dynamics of nature, where $T$ and $V$ are the kinetic and potential energy of the system. However, it should be kept in mind that this type of choice of the Lagrangian is nothing but a special choice \cite{Rana,Raychaudhuri,Goldstein}. There may exist several forms of the Lagrangian for which the Euler-Lagrange equation of motion is preserved. In this scenario, the general choice of the Lagrangian should be a non-canonical type \cite{Jirousek}. In addition, within the context of special relativistic dynamics, the classical idea of $\La=T-V$ is no longer deemed suitable \cite{Raychaudhuri}. Again, utilizing the non-standard Lagrangian in the action principle can potentially provide explanations for unresolved issues in the canonical sector. These include the matter-antimatter disparity, the cosmological constant problem, the dimensions and configuration of the universe, cosmic inflation, the horizon problem and other relevant aspects within the field of cosmology. The unification of gravity and quantum mechanics within a single theoretical framework is the biggest outstanding problem in the domain of fundamental physics. Thus, work still has to be done to explore more general platform as well as techniques. In the present context, therefore, we have chosen to employ the K-essence theory as a non-canonical theory for our investigation, which has been supported by the research of several authors \cite{Visser,Babichev1,Vikman,Chimento1,Picon1,Scherrer,Chimento2,Picon2,Picon3}.

In the K-essence theory \cite{Visser,Babichev1,Vikman,Chimento1,Picon1,Scherrer,Chimento2,Picon2,Picon3}, the scalar field is coupled minimally with gravity and involves in several types of non-canonical Lagrangian which can have the form of $\La(X,\phi)=-V(\phi)F(X)$ \cite{Vikman,Picon1,Picon2} or $\La (X,\phi)=F(X)-V(\phi)$ \cite{Dutta,Santiago} or $\La (X,\phi)\equiv \La (X)= F(X)$ \cite{Scherrer,Mukohyama} where $F(X)\equiv \La(X) (\neq X)$ is the non-canonical kinetic part, $F(X)$ is the non-canonical kinetic part, $X=\frac{1}{2} g_{\mu\nu} \nabla^{\mu}\phi \nabla^{\nu} \phi$, $V(\phi)$ is the canonical potential part. The use of the K-essence model by the authors of this research is employed in the investigation of primordial dark energy. Additionally, it is worth noting that there are instances of non-minimally coupled K-essence theories as discussed by Refs. \cite{Myrzakulov2,Sen,Chatterjee}. However, the focus of this article is solely on the minimally coupled K-essence theory, as explored by Refs. \cite{Visser,Babichev1,Vikman,Chimento1,Picon1,Scherrer,Chimento2,Picon2,Picon3}. Typically, the Lagrangian possesses the capability to rely on arbitrary functions of $\phi$ and $X$ in a broad sense. The advantage of the K-essence theory is that it can avoid the fine-tuning problem and produce the negative pressure responsible for the acceleration of the universe through the field's kinetic energy only. The potential term is suppressed by the kinetic term of the field. This may be a solution for the well-known Cosmic Coincidence problem \cite{Velten} that arises in the usual theory of general theory of relativity. In \cite{Picon1} we can find some attractor solutions in which the evolution of the universe is determined by the scalar field of the models. Because the K-essence field sub-dominated and imitated the radiation's equation of state, $\o$ (EoS), the ratio of the K-essence field to radiation density stayed constant during the radiation-dominated phase. Dynamical limitations prevented the K-essence field from reproducing the dust-like EoS during the dust-dominated period, but it quickly decreased its energy value by many orders of magnitude and achieved a constant value. Later, at a time roughly equal to the age of the universe today, the matter density was suppressed by the K-essence field and the cosmos began to accelerate. The K-essence theory's EoS eventually returns to a value between $0$ and $-1$. Though theoretically, it can go beyond $-1$. The author of \cite{Vikman1} showed that transitions from $\omega\geq -1$ to $\omega< -1$ (or vice versa) of dark energy, as represented by a generic scalar-field Lagrangian, are either unstable when subjected to cosmic perturbations or occur only on trajectories of negligible measure.   Even if the cosmos is mostly dominated by dark energy, this outcome remains strong even when other energy components interact with dark energy through non-kinetic couplings. The possibility of the K-essence theory to produce a form of dark energy in which sound travels at a constantly slower speed than light is another fascinating feature of the theory. The cosmic microwave background (CMB) disturbances on large angular scales may be lessened by this attribute \cite{Erickson,Dedeo,Bean}. The observational evidence for the K-essence theory and other modified theories have been mentioned in \cite{Planck1,Planck2}. 

In this particular context, Manna et al. \cite{gm1,gm2,gm3,gm4,gm5,gm6,gm7,gm8,gm9,gm10,gm11} have formulated an interesting emergent gravity metric, denoted as $\G_{\mu\nu}$. This metric exhibits additional characteristics compared to the conventional gravitational metric $g_{\mu\nu}$ and is derived based on the principles of the Dirac-Born-Infeld (DBI) type action \cite{Mukohyama,Born,Heisenberg,Dirac}. Dirac et al. \cite{Dirac} suggested a non-canonical Lagrangian to get rid of the infinite self-energy of the electron. Later, this type of Lagrangian found its vast use in the field of cosmology and quantum gravity \cite{Linde,Albrecht,Dvali,Kachru,Alishahiha,Silverstein,Chen1,Weinberg,Chen2}. In order to construct the $f(\R, \La(X))$ type modified theory of gravity and its cosmological implications, Manna et al. \cite{gm8} used the emergent metric described in \cite{gm1,gm2,gm3}. Manna \cite{gm6} and Ray et al. \cite{gm7} conducted research on the characteristics of singularity within the framework of the generalized Vaidya metric, employing the aforementioned non-canonical method in different settings. The researchers found that under certain precise circumstances, the singularity might manifest as a naked singularity. The research has also examined the strength of uniqueness in their study. Majumder et al. \cite{Majumder} recently investigated the radial and non-radial geodesic structures of the generalized K-essence Vaidya spacetime. Two forms of generalized K-essence Vaidya mass functions were utilized. They have suggested that wormholes may exist at the extreme phases of spacetime, especially in connection to black holes and white holes, similar to the Einstein-Rosen bridge. They have also observed a unique sign of quantum tunneling near the singularity. Panda et al. \cite{Panda} conducted a study on the $f(\R,\T)$ gravity theory, employing the emergent metric $(\G_{\mu\nu})$ inside a non-canonical geometry. Their findings indicate a strong agreement between the equation of state (EOS) parameter and the observational data. On the other end of the spectrum, the K-essence idea is commonly employed for the purpose of investigating dark energy as a model. However, it can be utilized solely from a gravitational or geometrical perspective \cite{gm5,gm6,gm7,gm9,gm10,gm11} 
due to the ongoing debate about the presence of dark energy \cite{Nielsen} based on current observations \cite{Ade}.

This study aims to investigate the fate of singularity within the framework of $f(\R,\T)$ gravity in generalized emergent Vaidya spacetime. To do this, we employ the non-canonical technique known as the K-essence. The field equation of $f(\R,\T)$ gravity, derived from emergent geometry, is studied in the present research~\cite{Panda}. The investigation focuses on the fate of the singularity by taking the emergent Vaidya metric as the primary metric~\cite{gm6}. In {Section} II, we provide a concise overview of the K-essence model in connection with the associated $f(\R, \T)$ gravity. The revised emergent field equation has a different geometric nature compared to the equation developed by Harko et al.~\cite{Harko2}. The solution for the emergent Vaidya metric within the framework of $f(\R, \T)$ gravity is presented in Section III. In this section, the major measure considered is the generalized emergent Vaidya metric. By analyzing the given information, we are able to derive the distinct components of the modified emergent field equation. We obtain the various mass functions for various choices of the function $f(\R, \T)$. In {Section} IV, we conducted an extensive study of the collapsing prospects of the given spacetime, considering various options of the function $f(\R, \T)$. In this analysis, a global naked singularity has been found, which is associated with the gravitational collapse under certain selections of the function $f(\R, \T)$. Conversely, our observations have shown examples of the universe exhibiting acceleration, mostly driven by the presence of dark energy. Additionally, we have detected the presence of both positive and negative masses, which give rise to gravitational dipoles. In {Section V, we have} formulated the strong or weak conditions of the singularity. The last {Section VI} is our conclusion of the work.

\section{Brief review of K-essence theory and the corresponding $f(\bar{R},\bar{T})$ gravity}
This section provides a concise overview of K-essence geometry and the associated $f(\bar{R},\bar{T})$ gravity. Firstly, we provide a concise overview of the geometry associated with the K-essence, as discussed in many academic sources \cite{Visser,Babichev1,Vikman,Chimento1,Picon1,Scherrer,Chimento2,Picon2,Picon3}. The action of this geometry is 
\ben
S_{k}[\phi,g_{\mu\nu}]= \int d^{4}x {\sqrt -g} \La(X,\phi)
\label{1}
\een
where $X=\frac{1}{2}g^{\mu\nu}\nabla_{\mu}\phi\nabla_{\nu}\phi$ is the canonical kinetic term and $\La(X,\phi)$ is the non-canonical Lagrangian. Here, the typical gravitational metric $g_{\mu\nu}$ has minimally coupled with the K-essence scalar field ($\phi$). 

The corresponding energy-momentum tensor associated with the K-essence scalar field only is \cite{Vikman}:
\ben
T_{\mu\nu}\equiv  \frac{-2}{\sqrt {-g}}\frac{\delta S_{k}}{\delta g^{\mu\nu}}=-2\frac{\partial \La}{\partial g^{\mu\nu}}+g_{\mu\nu}\La
=-\La_{X}\nabla_{\mu}\phi\nabla_{\nu}\phi
+g_{\mu\nu}\La,
\label{2}
\een
where $\La_{\mathrm X}= \frac{d\La}{dX},~ \La_{\mathrm XX}= \frac{d^{2}\La}{dX^{2}},
~\La_{\mathrm\phi}=\frac{d\La}{d\phi}$ and  $\nabla_{\mu}$ is the covariant derivative defined with respect to the gravitational metric $g_{\mu\nu}$. 

The K-essence scalar field equation of motion (EOM) is \cite{Babichev1,Vikman}
\ben
-\frac{1}{\sqrt {-g}}\frac{\delta S_{k}}{\delta \phi}= \tilde{G}^{\mu\nu}\nabla_{\mu}\nabla_{\nu}\phi +2X\La_{X\phi}-\La_{\phi}=0,
\label{3}
\een
where  
\ben
\tilde{G}^{\mu\nu}\equiv \frac{c_{s}}{\La_{X}^{2}}[\La_{X} g^{\mu\nu} + \La_{XX} \nabla ^{\mu}\phi\nabla^{\nu}\phi],
\label{4}
\een
with $1+ \frac{2X  \La_{XX}}{\La_{X}} > 0$ and $c_s^{2}(X,\phi)\equiv{(1+2X\frac{\La_{XX}}
{\La_{X}})^{-1}}$.

The inverse metric is \cite{Babichev1,Vikman}
\ben G_{\mu\nu}=\frac{\La_{X}}{c_{s}}[g_{\mu\nu}-{c_{s}^{2}}\frac{\La_{XX}}{\La_{X}}\nabla_{\mu}\phi\nabla_{\nu}\phi].
\label{5}
\een

After a conformal transformation \cite{gm1,gm2} $\bar G_{\mu\nu}\equiv \frac{c_{s}}{\La_{X}}G_{\mu\nu}$ we have
\ben
\bar{G}_{\mu\nu}=g_{\mu\nu}-\frac{\La_{XX}}{\La_{X}+2X\La_{XX}}\nabla_{\mu}\phi\nabla_{\nu}\phi.
\label{6}
\een

The Eqs. (\ref{4})--(\ref{6}) hold physical significance under the condition that $\La_{X}$ is not equal to zero, given a positive definite $c_{s}^{2}$. Eq. (\ref{6}) asserts that the emergent metric, denoted as $\bar{G}_{\mu\nu}$, is conformally different from the metric $g _{\mu\nu}$ when considering non-trivial configurations of the scalar field $\phi$. Similar to canonical scalar fields, the variable $\phi$ has varied local causal structural characteristics. It is also different from those that are defined with $g_{\mu\nu}$. The equation of motion, as expressed in Eq. (\ref{3}), remains applicable even when considering the implicit dependence of $\La$ on $\phi$. Then the EOM Eq. (\ref{3}) is:
\ben
\frac{1}{\sqrt{-g}}\frac{\delta S_{k}}{\delta \phi}= \bar G^{\mu\nu}\nabla_{\mu}\nabla_{\nu}\phi=0.
\label{7}
\een

In this study, we look into the Dirac-Born-Infeld (DBI) type non-canonical Lagrangian denoted as $\La(X,\phi)\equiv \La(X)$ \cite{gm1,gm2,gm3,Mukohyama,Born,Heisenberg,Dirac,Panda}:
\ben
\La(X)= 1-\sqrt{1-2X}.
\label{8}
\een

The K-essence framework argues that the dominance of kinetic energy over potential energy leads to the omission of the potential term in the Lagrangian equation (\ref{8})~\cite{Mukohyama,Panda}. Consequently, the squared speed of sound, denoted as $c_{s}^{2}$, is given by $(1-2X)$. Thus, the {\it effective emergent metric} Eq. (\ref{6}) becomes
\ben
\bar G_{\mu\nu}= g_{\mu\nu} - \nabla_{\mu}\phi\nabla_{\nu}\phi= g_{\mu\nu} - \partial_{\mu}\phi\partial_{\nu}\phi,
\label{9}
\een
since $\phi$ is a scalar. 

Following~\cite{gm1,gm2}, the Christoffel symbol associated with the emergent gravity metric Eq. (\ref{9}) is: 
\ben
\bar\Gamma ^{\alpha}_{\mu\nu} 
=\Gamma ^{\alpha}_{\mu\nu} -\frac {1}{2(1-2X)}\Big[\delta^{\alpha}_{\mu}\partial_{\nu}
+ \delta^{\alpha}_{\nu}\partial_{\mu}\Big]X,~~~~~~~~~~~
\label{10}
\een
where $\Gamma ^{\alpha}_{\mu\nu}$ is the usual Christoffel symbol associated with the gravitational metric $g_{\mu\nu}$.

Therefore, the geodesic equation for the K-essence geometry becomes:
\ben
\frac {d^{2}x^{\alpha}}{d\l^{2}} +  \bar\Gamma ^{\alpha}_{\mu\nu}\frac {dx^{\mu}}{d\l}\frac {dx^{\nu}}{d\l}=0, \label{11}
\een
where $\l$ is an affine parameter.

The covariant derivative $D_{\mu}$ \cite{Babichev1} linked with the emergent metric $\bar{G}_ {\mu\nu}$ $(D_{\a}\bar{G}^{\a\b}=0)$ yields
\ben
D_{\mu}A_{\nu}=\partial_{\mu} A_{\nu}-\bar \Gamma^{\l}_{\mu\nu}A_{\l}, \label{12}
\een
and the inverse emergent metric is $\bar G^{\mu\nu}$ such as $\bar G_{\mu\l}\bar G^{\l\nu}=\delta^{\nu}_{\mu}$.

Hence, when taking into account the comprehensive action that characterizes the dynamics of K-essence and general relativity \cite{Vikman}, the Emergent Einstein's Equation (EEE) may be expressed as follows:
\ben
\mathcal{\G}_{\mu\nu}=\R_{\mu\nu}-\frac{1}{2}\bar{G}_{\mu\nu}\R=\k \T_{\mu\nu}, \label{13}
\een
where $\k=8\pi G$ is constant, $\R_{\mu\nu}$ is Ricci tensor and $\R~ (=\R_{\mu\nu}\bar{G}^{\mu\nu})$ is the Ricci scalar. Additionally, the energy-momentum tensor $\T_{\mu\nu}$  is associated with this emergent spacetime. In order to establish a comprehensive understanding of the energy-momentum tensor associated with this particular geometry, it is important to proceed with its definition as~\cite{Panda}
\ben
\T_{\mu\nu}=-\frac{2}{\sqrt{-\G}}\frac{\partial\Big(\sqrt{-\G}\La(X)\Big)}{\partial \G_{\mu\nu}},
\label{14}
\een
where $\big(-\G \big)^{1/2}=\big(-det({\G_{\mu\nu}})\big)^{1/2}$.

Now, we will provide a concise overview of the $f(\R,\T)$ gravity theory within the framework of K-essence geometry, drawing inspiration from the work of Harko et al.~\cite{Harko2}. The phenomenon of modified gravity within the context of K-essence geometry has been discussed in a study by Panda et al. \cite{Panda}.

The action ($\k=1$) of this type of modified gravity is \cite{Panda}:
\ben
S=\int d^4x\sqrt{-\G}\Big[f(\R,\T)+\La(X)\Big],
\label{15}
\een
where the function $f(\R,\T)$ represents an arbitrary function of the Ricci scalar ($\R$) and the trace of the energy-momentum tensor ($\T=\T^{\mu\nu}\G_{\mu\nu}$). On the other hand, $\La(X)$ denotes the non-canonical Lagrangian associated with the K-essence geometry. 

The updated action (\ref{15}) reveals a clear dependence on the variables $\R$, $\T$, and $X(=\frac{1}{2}g^{\mu\nu}\nabla_{\mu}\phi\nabla_{\nu}\phi)$, rather than on the explicit dependence of the K-essence scalar field ($\phi$). Considering $\bar{G}_{\mu\nu}$ as in Eq. (\ref{9}), the emergent energy-momentum tensor (\ref{14}) can also be written as 
\ben
\T_{\mu\nu}
=\G_{\mu\nu}\La(X)-2\frac{\partial \La(X)}{\partial \G^{\mu\nu}}.
\label{16}
\een

Varying the action and following \cite{Panda} we get {\it the modified field equation of emergent $f(\R,\T)$ gravity} for $\La(X)$ taken as Eq. (\ref{8})
\ben
F\R_{\mu\nu}-\frac{1}{2}f(\R,\T)\G_{\mu\nu}+(\G_{\mu\nu}\bar{\square}-D_{\mu}D_{\nu})F=\frac{1}{2}\T_{\mu\nu}-f_{\T}\T_{\mu\nu}-f_{\T}\bar{\Theta}_{\mu\nu}.
\label{17}
\een
where
\ben
\bar{\Theta}_{\mu\nu}=\G^{\alpha\beta}\frac{\partial \T_{\alpha\beta}}{\partial\G^{\mu\nu}}
=\G_{\mu\nu}\La(X)-2\T_{\mu\nu}-2\G^{\a\b}\frac{\partial^2 \La(X)}{\partial \G^{\mu\nu}\partial \G^{\a\b}},
\label{18}
\een 
with $\bar{\square}=D_{\mu}D^{\mu}$, $F(\R,\T)=\partial f(\R,\T)/\partial \R\equiv F$ and  $f_{\T}(\R,\T)=\partial f(\R,\T)/\partial \T\equiv f_{\T}$ respectively. Since the background K-essence geometry differs from the ordinary gravity, the modified field equation (\ref{17}) of emergent $f(\R, \T)$ gravity \cite{Panda} is completely different from the usual $f(R,T)$ gravity produced by Harko et al.~\cite{Harko2} in their work.

\section{Solution of emergent Vaidya Metric in $f(\bar{R},\bar{T})$ gravity}

The purpose of this section is to investigate the emergent Vaidya metric in the context of $f(\R,\T)$ gravity and evaluate its characteristics. In this study, we focus on the primary metric known as the emergent Vaidya type, as introduced by Manna et al. in their work~\cite{gm6}, utilizing the K-essence geometry. The K-essence emergent Vaidya line element is 
\ben
dS^2=-\Big[1-\frac{2\M(v,r)}{r}\Big]dv^2+2dvdr+r^2d\Omega^2,
\label{19}
\een
where $\M$ is the K-essence emergent Vaidya mass function having the form
\ben
\M(v,r)=m(v,r)+\frac{r}{2}\phi_v^2,
\label{20}
\een
where $d\Omega^2=d\theta^2+\sin^{2}\theta~d\Phi^2$, $\phi_v=\frac{\p\phi}{\p v}$ and $\phi_v^2$ denote the non-zero kinetic energy associated with the K-essence scalar field. 

The mass function $m(v,r)$ is referred to as the mass function for the conventional generalized Vaidya spacetime, which serves as the underlying gravitational metric. Alternatively, the function $m(v,r)$ represents the mass distribution associated with the gravitational energy within a certain radius $r$. The variable $v$ corresponds to the null coordinate that corresponds to the Eddington advanced time, while $r$ decreases towards the future along a ray defined by $v=\text{constant}$~\cite{Husain,Wang,Mkenyeleye}. According to the metric (\ref{19}), the values of $\phi_v^{2}$ are between $0$ and $1$, as well as a non-decreasing function of $v$ under the energy conditions \cite{gm5}. This is a somewhat significant constraint on the types of K-essence scalar field configurations that can be used in the emergent generalized Vaidya solution. The authors in the cited work \cite{gm6} have employed the emergent gravity metric, as shown by Eq. (\ref{9}), in conjunction with the Lagrangian of the DBI type, as represented by Eq. (\ref{8}).  In this context, it is crucial to note that we use a homogeneous K-essence scalar field indicated by $\phi(v,r)\equiv \phi(v)$ \cite{gm5,gm6,Panda,Panda1}. Because the dynamical solutions of the K-essence scalar fields result in the spontaneous violation of Lorentz symmetry, the possibilities given above are appropriate. Furthermore, it alters the metric pertaining to the perturbation in the vicinity of these solutions~\cite{Babichev1}.

The investigation carried out by Manna et al.~\cite{gm6} included the calculation of the emergent energy-momentum tensor ($\T_{\mu\nu}$).  This calculation took into account the specific form proposed by Husain et al.~\cite {Husain}, Mkenyeleye et al.~\cite{Mkenyeleye} and Wang et al.~\cite{Wang}  which states that 
\ben
{\T}_{\mu\nu}={\T}_{\mu\nu}^{(n)} + {\T}_{\mu\nu}^{(m)},
\label{21}
\een
with $\T_{\mu\nu}^{(n)}=\bar{\gamma} l_{\mu} l_{\nu}$ being the contribution due to Vaidya null radiation and $\T_{\mu\nu}^{(m)}=(\bar{\rho}+\bar{p})(l_{\mu}n_{\nu}+l_{\nu}n_{\nu})+\G_{\mu\nu}\bar{p}$, where $\bar{\rho}$ and $\bar{p}$ are the energy density and pressure of the emergent perfect fluid, respectively. The vectors $l_{\mu}$ and $n_{\mu}$ are two null vectors. The symbol $\bar{\gamma}$ represents the energy density associated with the emergent Vaidya null radiation. It is important to acknowledge that the emergent energy-momentum tensor (\ref{16}) is not dependent on the scalar field $\phi$, but rather on its time derivative. This is the reason why the vorticity term is not present in this emergent energy-momentum tensor. Hence, it can be observed that in the framework of the K-essence model, the use of the DBI type non-canonical Lagrangian (\ref{8}) has a resemblance to ideal fluid models without vorticity \cite{Vikman,Babichev1}. In this setting, we have employed the energy-momentum tensor $(\T_{\mu\nu})$ in a way consistent with a perfect fluid. These null vectors are defined as \cite{Rudra3}: $l_{\mu}=(1,0,0,0)$, $n_{\mu}=(\frac{1}{2}(1-\frac{\M(v,r)}{r}),-1,0,0)$ with $l_{\mu}l^{\mu}=n_{\mu}n^{\mu}=0$ and $l_{\mu}n^{\mu}=-1$. With these definitions of null vectors, we can write the emergent energy-momentum tensor similarly, as~\cite{Rudra3}
\ben
&&\T_{00}=\Big[\bar{\gamma}+\bar{\rho}(1-\frac{2\M(v,r)}{r})\Big],\nonumber\\
&&\T_{01}\equiv\T_{10}=-\bar{\rho}~,~\T_{22}=\bar{p}r^{2},\nonumber\\
&&\T_{33}=\bar{p}r^{2}\sin^{2}\theta,
\label{22}
\een
where the corresponding values of $\bar{\gamma}$, $\bar{\rho}$ and $\bar{p}$ are ($\kappa=1$): 
$\bar{\gamma}=\frac{2\M_v}{ r^2}=\frac{2m_v}{ r^2}+\frac{2\phi_v\phi_{vv}}{ r}$,~ 
$\bar{\rho}=\frac{2\M_r}{r^2}=\frac{2m_r}{ r^2}+\frac{\phi_v}{ r^2}$,~ 
and $\bar{p}=-\frac{\M_{rr}}{ r}=-\frac{m_{rr}}{ r}$. The corresponding energy conditions \cite{Hawking} for the energy-momentum tensor ${\T}_{\mu\nu}$ combination of Type-I and Type-II fluids can be found in \cite{gm6}:
(a) The weak and strong energy conditions $\bar{\gamma}\geq0~,~\bar{\rho}\geq0~,~\bar{p}\geq0~~~(\bar{\gamma}\neq0)$ and 
(b) The dominant energy conditions
$\bar{\gamma}\geq0~,~\bar{\rho}\geq\bar{p}\geq0  ~~~~~(\bar{\gamma}\neq0)$ provided that the conditions imposed on ${\T}_{\mu \nu}$ as
$\bar{\gamma} > 0 \Rightarrow  m_{v}+r\phi_v \phi_{vv} > 0$;~~~$\bar{\rho} > 0  \Rightarrow  2m_{r}+\phi_{v}^{2}>0$;~~~
$\bar{p}>0 \Rightarrow  m_{rr} < 0$.

To explore the phenomenon of gravitational collapse in the framework of the non-canonical $f(\R,\T)$ theory of gravity, we are now focusing on the emergent line element (\ref{19}) with the emergent Vaidya mass function (\ref{20}). Additionally, we want to determine the ultimate outcome of the singularity. Here, we utilize the suitable energy-momentum tensor (\ref{22}) to solve the modified field Eq. (\ref{17}) together with the Eq. (\ref{18}). For the purpose of this study, it is imperative to choose a particular form of the function $f(\R,\T)$ \cite{Rudra3}, which can be represented as $f(\R, \T)=f_{1}(\R)+f_{2}(\T)$. Although there may be alternative options, we cannot include them in the present study to maintain simplicity~\cite{Harko2,Mishra1}. The choice of these functions will vary depending on the specific kinds of $f_{1}(\R)$ and $f_{2}(\T)$ being considered. Then the modified field Eq. (\ref{17}) becomes
\ben
f_{1}'(\R)\R_{\mu\nu}-\frac{1}{2}f_{1}(\R)\bar{G}_{\mu\nu}+\Big(\bar{G}_{\mu\nu}\bar{\square}-D_{\mu}D_{\nu}\Big)f_{1}'(\R)=\frac{1}{2}\T_{\mu\nu}+\frac{1}{2}f_{2}(\T)\bar{G}_{\mu\nu}
-f_{2}'(\T)\Big[\T_{\mu\nu}+\bar{\Theta}_{\mu\nu}\Big],
\label{23}
\een
where $f_{1}'(\R)=\frac{\p f(\R,\T)}{\p\R}\equiv \frac{\p f_{1}(\R)}{\p \R}$ and $f_{2}'(\T)=\frac{\p f(\R,\T)}{\p\T}\equiv \frac{\p f_{2}(\T)}{\p \T}$. 

The non-vanishing components of the emergent field Eq. (\ref{23}) can be found as:\\
(i) The $(00)$ component is:
\ben
&-\bar{\gamma}+\frac{\Big(f_{1}(\R)+f_{2}(\T)\Big)\Big(r-2\M(v,r)\Big)}{2r}+\bar{\rho} \Big(\frac{2\M(v,r)}{r}-1\Big)+f_{2}'(\T)\Bigg[\bar{\gamma}+\bar{\rho}-\frac{2\bar{\rho}\M(v,r)}{r}-2\Big(\bar{\gamma}+\bar{\rho}-\frac{2\bar{\rho}\M(v,r)}{r}\Big)
+\nonumber\\
&\frac{\phi_v^4\Big(-5-8\phi_v^2+4\phi_v^4\Big)}{2\sqrt{1-\phi_v^2}}+\frac{\Big(r-2\M(v,r)\Big)\Big(-1+\sqrt{1-\phi_v^2}\Big)}{r}\Bigg]+f_{1}'(\R)\frac{2\dot{\M}(v,r)-(r-2\M(v,r))\M''(v,r)}{r^2}=0,
\label{24}
\een
(ii) the $(01)$ component is:
\ben
&-r\Big[f_1(\R)+f_2(\T)-2\bar{\rho} \Big(1+f_2'(\T)\Big)
-2f_2'(\T)\Big(1-\sqrt{1-\phi_v^2}\Big)\Big]+2f_{1}'(\R)\M''(v,r)=0,~~~~~~
\label{25}
\een
(iii) the $(22)$ or $(33)$ component of the field equation is:
\ben
r^2\Big[f_{1}(\R)+f_{2}(\T)+2\omega\bar{\rho} -2f_{2}'(\T)\Big(-1+\bar{\rho}\omega+\sqrt{1-\phi_v^2}\Big)\Big]-4f_{1}'(\R)\M'(v,r)=0.~~~~
\label{26}
\een

It should be mentioned that the $(00)$ component of the emergent field Eq. (\ref{23}) is not only too large it also contains both the derivatives of $\M(v,r)$ with respect to $v$ and $r$. The solution of this specific component of the field equation is a challenging task. We will now focus on the remaining two components of the emergent field Eq. (\ref{23}), namely the (01) and (22) or (33) components, designated as  Eqs. (\ref{25}) and (\ref{26}) respectively.

Now we look into several forms of $f_{1}(\R)$ and $f_{2}(\T)$ in the following manner:\\

{\bf Case 1:}
First, we consider the form of $f(\R, \T)$ \cite{Rudra3} as
 \ben
f(\R, \T)=f_{1}(\R)+f_{2}(\T)=g_1\R^{\b_1}+g_2 \T^{\b_2},
 \label{27}
\een
where $g_{1},~\b_{1},~g_{2},~\b_{2}$ are constants. Note that for this choice of $f(\R, \T)$ (\ref{27}), the field Eq. (\ref{23}) can only be solved for the choices of $\b$ as $\b_1=\b_2=1$. 

Using the generalized emergent Vaidya line element (\ref{19}) with the emergent Vaidya mass function (\ref{20}) and the emergent energy-momentum tensor (\ref{22}), we evaluate the $(22)$ or $(33)$ component of the field equation (\ref{26}) as:
\ben
g_1 \M''(v,r)+n r \Big(g_2 (2 \omega -1)+\omega \Big) \M(v,r)+g_2 r \Big(\sqrt{1-\phi_v^2}-1\Big)=0,
\label{28}
\een
where $\omega$ is the EoS parameter given by the relation $\bar{p}=\omega \bar{\rho}$ and the density can be written in terms of the mass parameter as $\bar{\rho}=n\M(v,r)$, $n(>0)$ being the particle number density~\cite{Rudra3,Boer} and $\M'(v,r)=\frac{\p\M(v,r)}{\p r}$,~ $\M''(v,r)=\frac{\p^{2}\M(v,r)}{\p r^{2}}$. Mention that the relation $\bar{p}=\omega \bar{\rho}$ may be used to get the values of pressure $(\bar{p})$ and energy density $(\bar{\rho})$. The above Eq. (\ref{28}) can also be written in a simple form as:
\ben
\M''(v,r)+\frac{A}{g_1}\M(v,r) r=k r,
\label{29}
\een
where $A=n\Big(w+g_2(2w-1)\Big)$ and $k=\frac{g_{2}}{g_{1}}\Big(1-\sqrt{1-\phi_v^2}\Big)$.

The solution of the above differential Eq. (\ref{29}) is given by
\ben
\M(v,r)=
&\frac{1}{A}\Big[\pi g_1 k AiryAi\Big(-\frac{A r}{(-\frac{A}{g_1})^{2/3} g_1}\Big) AiryBi'\Big(\sqrt[3]{-\frac{A}{g_1}} r\Big)-\pi g_1 k AiryAi'\Big(\sqrt[3]{-\frac{A}{g_1}} r\Big) AiryBi\Big(-\frac{A r}{(-\frac{A}{g_1})^{2/3} g_1}\Big)\Big]\nonumber\\
&+c_1(v) AiryAi\Big(-\frac{A r}{(-\frac{A}{g_1})^{2/3} g_1}\Big)+c_2(v) AiryBi \Big(-\frac{A r}{(-\frac{A}{g_1})^{2/3} g_1}\Big),
\label{30}
\een
where $AiryAi$, $AiryAi'$, $AiryBi$ and $AiryBi'$  are the four Airy functions defined in \cite{Rudra3,Lebedev} and the functions $c_1(v)$ and $c_2(v)$ are arbitrary functions of time that result from the process of integration. It is important to acknowledge that the Airy function is a special mathematical function that has been called in honour of the renowned British astronomer George Biddell Airy (1801-1892). There are two distinct Airy functions, denoted as $Ai(x)$ (Airy function of the first kind) and $Bi(x)$ (Airy function of the second kind). These functions are considered to be linearly independent solutions to the Airy differential equation, which may be expressed as $\frac{d^{2}y}{dx^{2}}-xy=0$.

The $(01)$ component (\ref{25}) of the field Eq. (\ref{23}) can be expressed as:
\ben&2 g_1 \Big(2 \M'(v,r)+r \M''(v,r)\Big)+2 r^2 g_2 n (\omega -1) \M(v,r)=0
\label{31}
\een
and the solution of this Eq. (\ref{31}) becomes (\ref{23}) can be expressed as:
\ben
\M(v,r)=\frac{1}{r}\Bigg[c_3(v) AiryAi\Big(\frac{r (g_2 n-g_2 n \omega )}{g_2 \Big(\frac{g_2 n-g_2 n \omega }{g_1}\Big)^{2/3}}\Big)+c_4(v) AiryBi\Big(\frac{r (g_2 n-g_2 n \omega )}{g_1\Big(\frac{g_2 n-g_2 n \omega }{g_1}\Big)^{2/3}}\Big)\Bigg].
\label{32}
\een

It is evident that between these two solutions the first solution (\ref{30}) is much more general than the second one (\ref{32}). It has also been verified that for some particular choices of initial conditions, we can get back Eq. (\ref{32}) from Eq. (\ref{30}). Most importantly the contributions of the K-essence scalar field are absent in the second solution. Therefore, we have decided to work with the first solution of the field equation to check the fate of singularity. It should also be highlighted that for every other option, we solve just Eq. (\ref{26}) for the same reasons indicated above.\\

{\bf Case 2:}
The form $f(\R,\T)$ in this case is
\ben
f(\R,\T)=f_1 (\R)+f_2 (\T)=g_1 \R^{\b_1}+g_2 e^{\b_2 \T}.
\label{33}
\een

The $(22)$ and $(33)$ component of the field equation becomes
\ben
&-\frac{1}{2} r^2 \Big(g_1 \Big(\frac{4 \M'(v,r)+2 r \M''(v,r)}{r^2}\Big)^{\beta _1}+g_2 e^{2 \beta _2 n (\omega -1) \M(v,r)}\Big)+2 \beta _1 g_1 \M'(v,r) \Big(\frac{4 \M'(v,r)+2 r \M''(v,r)}{r^2}\Big)^{\beta _1-1}\nonumber\\
&-\beta _2 g_2 r^2 e^{2 \beta _2 n (\omega -1) \M(v,r)} \Big(n \omega  \M(v,r)+\sqrt{1-\phi_v ^2}-1\Big)-n r^2 \omega  \M(v,r)=0.
\label{34}
\een

The solution for $\M(v,r)$ can be achieved for the condition $\b_1=1$ and $\b_2=0$. Hence, we get
\ben
&\M(v,r)=
\frac{1}{2n\omega}\Bigg[AiryAi\Big(r \sqrt[3]{-\frac{n \omega }{g_1}}\Big) \Big(-\pi  g_2 AiryBi'\Big(r \sqrt[3]{-\frac{n \omega }{g_1}}\Big)+2 d_1(v) n \omega \Big)\nonumber\\
&+\pi  g_2 AiryAi'\Big(r \sqrt[3]{-\frac{n \omega }{g_1}}\Big) AiryBi\Big(r \sqrt[3]{-\frac{n \omega }{g_1}}\Big)+2 d_2(v) n \omega  AiryBi\Big(r \sqrt[3]{-\frac{n \omega }{g_1}}\Big)\Bigg].
\label{35}
\een

In this specific case, the mass function $\M(v,r)$ also exhibits dependence on Airy functions as well as arbitrary functions of time, namely $d_1(v)$ and $d_2(v)$, which arise through the process of integration. Also, it is important to note that in the specific case when $\b_1=1$ and $\b_2=0$, Eq. (\ref{33}) has been reformulated as $f(\R,\T)=g_{1}\R+g_{2}$. For this specific kind of choice the $f(\R,\T)$ gravity is transformed into modified Einstein-Hilbert gravity by the action within the K-essence geometry and Eq. (\ref{35}) is a valid solution of mass function for $f(\R,\T)$ gravity. Note that $g_{2}$ should not be considered as the cosmological constant. \\

{\bf Case 3:}
In this case we take the function $f(\R,\T)$ as
\ben
f(\R, \T)=f_{1}(\R)+f_{2}(\T)=g_1e^{\b_1\R}+g_2 e^{\b_2\T}
\label{36}
\een
where $g_1,~g_2,~\b_1,~\text{and}~\b_2$ are constants. This special type of choice is often known as, the double-exponential (DE) model. The $(22)$ and $(33)$ component of the field equation (\ref{26}) becomes
\ben
&-\frac{1}{2} r^2 \Big[g_1 e^{\frac{1}{r^2}\Big(2 \beta _1 (2 \M'(v,r)+r \M''(v,r))\Big)}+g_2 \Big(2 \beta _2 (\sqrt{1-\phi_v ^2}-1)+1\Big) e^{2 \beta _2 n (\omega -1) \M(v,r)}\Big]\nonumber\\
&+2 \beta _1 g_1 \M'(v,r) \e^{\frac{1}{r^2} \Big(2 \beta _1 (2 \M'(v,r)+r \M''(v,r))\Big)}-n r^2 \omega  \M(v,r) \Big(\beta _2 g_2 e^{2 \beta _2 n (\omega -1) \M(v,r)}+1\Big)=0.
\label{37}
\een

We have exponential terms that include the first and second-order derivatives of $\M(v,r)$. So rather than finding a general solution, we look for an approximate solution, taking the consideration $\b_1=0$ and expanding the second and fourth exponential in Taylor's series. Other options for the aforementioned constants, easily verified, cannot be solved. Thus we obtain the solution of $\M(v,r)$ as

\ben
\M(v,r)=\frac{g_1\Big[-1+2\b_2(1+\sqrt{1-\phi_v^2})\Big]+g_2\Big[-1+2\b_2(1-2n+\sqrt{1-\phi_v^2})\Big]}{2g_2\b_2 n(\omega-1)\Big[1+\b_2\Big(n-2(1+\sqrt{1-\phi_v^2})\Big)\Big]}.
\label{38}
\een

The solution of $\M(v,r)$ no longer depends on $r$. However, there is an explicit dependence on $v$ through $\phi_v^2$, the values of this kinetic part of the K-essence field range between $0$ and $1$. Consequently, the collapse is not feasible in this particular case, as we shall elaborate on in the next Section.\\

{\bf Case 4:} In this case we choose the function $f(\R,\T)$ as 
\ben
f(\R,\T)=f_1(\R)+f_2(\T)=g_1 e^{\b_1 \R}+g_2 \T^{\b_2}.
\label{39}
\een

The $(22)$ or $(33)$ component of the field Eq. (\ref{26}) is:
\ben
&&-\frac{1}{2} r^2 \Big(g_1 \exp \Big(\frac{2 \beta _1 \Big(2 \M'(v,r)+r \M''(v,r)\Big)}{r^2}\Big)+2^{\beta _2} g_2 (n (\omega -1) \M(v,r))^{\beta _2}\Big)\nonumber\\
&&+2 \beta _1 g_1 \M'(v,r) \exp \Big(\frac{2 \beta _1 \Big(2 \M'(v,r)+r \M''(v,r)\Big)}{r^2}\Big)-2^{\beta _2-1} \beta _2 g_2 r^2 (n (\omega -1) \M(v,r))^{\beta _2-1} \Big(n \omega  \M(v,r)\nonumber\\
&&+\sqrt{1-\phi_v ^2}-1\Big)-n r^2 \omega  \M(v,r)=0.
\label{40}
\een

The solvability of $\M(v,r)$ is determined by the choice of the constants $\b_1=0$ and $\b_2=1$. It should be noted that other choices of constants do not yield solvable solutions.

Now
\ben
\M(v,r)=\frac{-2 g_2 \sqrt{1-\phi_v^2}-g_1+2 g_2}{2 n \Big(2 g_2 \omega -g_2+\omega \Big)}.
\label{41}
\een

The result bears similarity to Case 3, as the mass function $M(v,r)$ remains unaffected by the radial component but is dependent upon time due to the presence of the kinetic portion of the K-essence scalar field.

\section{Gravitational Collapse}

To study the type of singularity formed in our case we follow the approach of \cite{Rudra3,Heydarzade1,Heydarzade2,Mkenyeleye,gm6,gm7}. The authors~\cite{Rudra3,Khan,Abbas}, studied gravitational collapse through the study of the radial null geodesic emerging from the singularity in the case of usual $f(R, T)$ gravity. In this analysis, we will look at a system undergoing spherical collapse. Specifically, we will focus on the physical radius, denoted as $\mathcal{R}(v,r)$, of the $r-th$ shell of the star at time $v$. An appropriate starting condition can be defined as follows: at $v=0$, $\mathcal{R}(0,r)=r$. It is evident that in the case of an inhomogeneous collapse, distinct collapsing shells may reach singularity at varying points in time. Our focus lies on the light particles, known as photons, that emanate from the singularity and traverse through the geodesics, ultimately reaching an observer situated outside the singularity. The presence of an event horizon poses a barrier to the passage of photons, hence preventing their arrival to the observer. In this study, we will investigate the presence of outgoing non-spacelike geodesics. Theoretically, if such geodesics possess well-defined tangent at the singularity, the quantity $\frac{d\mathcal{R}}{dr}$ will definitely tend towards a finite limit with the geodesics approaching the singularity in the past following the trajectories. At the instances when these trajectories intersect the locations $(v_{0}, r)=(v_{0}, 0)$, an extensive breakdown in mathematical and physical principles takes place, leading to the emergence of a singularity at $\mathcal{R}(v_0,0)=0$. This differs from the scenario observed in regular black holes \cite{Tolman, Hayward1}. Ideally, at these specific locations, the matter shells undergo a process of collapse, ultimately reaching a radius of zero. This collapse leads to the emergence of a central singularity. This is a very compact object because a large quantity of mass is packed into a very small volume. The equation for outgoing radial null geodesic from the emergent Vaidya metric (\ref{19}) taking $dS^2=0$ and $d\Omega^2=0$ as (using Eq. (\ref{20})) \cite{gm6}
\ben
\frac{dv}{dr}=\frac{2}{\Big(1-\frac{2 \M(v,r)}{r}\Big)}=\frac{2}{\Big(1-\frac{2m(v,r)}{r}+\phi_v^2\Big)}.
\label{42}
\een

It is important to mention that the total mass function ($\M(v,r)$) must be separated into two components (\ref{20}), {\it viz.} one arising from the generalised Vaidya mass $(m(v,r))$ of the background metric and another arising from the kinetic portion ($\phi_{v}^{2}$) of the K-essence scalar field. In this K-essence Vaidya geometry, the scalar field is minimally coupled with the typical generalized Vaidya spacetime. Here, the scalar field $\phi(v,r)$ is equivalent to $\phi(v)$, indicating that it is just a homogeneous K-essence scalar field. Clearly, the differential equation (\ref{42}) exhibits a singularity at the values of $r = 0$ and $v = 0$. Mathematically, this implies that any solution to the equation mentioned above lacks analyticity at the point when $r = 0$ and $v = 0$. Due to the mathematical breakdown occurring at the singularity, it is necessary to analyze the behavior as one approaches to the singularity. To do this let us introduce a parameter, $Y=v/r$. Now we will examine the nature of $Y$ at the limiting point,  $r\rightarrow 0$, $v\rightarrow 0$. Let us consider at  $r\rightarrow 0$, $v\rightarrow 0$ the value of $Y$ is $Y_0$. Then we can write \cite{Rudra3,gm6}
\ben
Y_0&=&\lim \limits_{\substack{%
    v \to 0\\
    r \to 0}}Y=\lim \limits_{\substack{%
    v \to 0\\
    r \to 0}} \frac{v}{r} =\lim \limits_{\substack{%
    v \to 0\\
    r \to 0}} \frac{dv}{dr}=\lim \limits_{\substack{%
    v \to 0\\
    r \to 0}} \frac{2}{1-(\frac{2\M(v,r)}{r})}\nonumber\\
    &=&\lim \limits_{\substack{%
    v \to 0\\
    r \to 0}} \frac{2}{\Big(1-\frac{2m(v,r)}{r}+\phi_v^2\Big)}\equiv b_{\pm}.
\label{43}
\een

This will provide an algebraic equation expressed in relation to $Y_0$ \cite{Rudra3,gm6,Mkenyeleye}, with $b_{\pm}$ representing the positive and negative roots of $Y_{0}$, respectively. Our primary focus will be on determining the roots of this Eq. (\ref{43}), since they directly correspond to the slopes (directions) of the tangents to the geodesics. In this context, we are just concerned with the actual solutions to the problem, since we are focusing on a practical situation that does not have any link to the complex domain. In our configuration, any positive real solution of this algebraic equation will determine the direction of the tangent to an outgoing null geodesic at the singularity. Thus, the presence of positive real roots of this equation indicates a necessary and sufficient condition for the singularity to be naked in nature, whereas the negativity of the roots represents the formation of the Black Hole in the future~\cite{Heydarzade2,gm6}. We know that a ray released from the singularity would reach the outside observer if a single null geodesic in the $(v, r)$ plane escapes from it and the observer would be able to observe the singularity, which makes the singularity {\it locally naked} for an instant of time. Observing the singularity by the observer such that the observer extracts some information is necessary for a bunch of geodesics to escape from the $(v,r)$ plane. In other words, this phenomenon excludes the existence of an event horizon. The escaping of the family of null geodesic makes a {\it globally naked singularity} \cite{Joshi3,Joshi4,Joshi5}. The comprehensive explanation and interpretation of $Y_0$, together with the associated emergent mass function (\ref{20}), its derivatives and the K-essence scalar field as well as its derivatives at the central singularity $(r\rightarrow 0,~ v\rightarrow 0)$, may be found in the Ref. \cite{gm6}. It is essential to keep in mind that in this particular section, we undertake an investigation of the characteristics of various mass functions. These functions are obtained from the various choices of $f(\R,\T)$ within the framework of K-essence geometry, specifically in connection to the emergent Vaidya spacetime.\\

{\bf Case-1}:
Following~\cite{Rudra3}, using Eq. (\ref{30}) in Eq. (\ref{43}) at the central singularity and considering the value of $AiryAi(0)$, $AiryBi(0)$, $AiryAi'(0)$ and $AiryBi'(0)$ we get the equation of $Y_0$ after some algebraic calculations
\ben
Y_0^2\Bigg[\frac{\xi_1}{3^{2/3}\Gamma(2/3)}+\frac{\xi_2}{3^{1/6}\Gamma(2/3)}\Bigg]-Y_0\Bigg[\frac{\phi_v^4}{4}+\frac{1}{2}\Bigg]+1=0,
\label{44}
\een
where the integration constants $c_1(v)$ and $c_2(v)$ in Eq.(\ref{30}) are chosen as $c_1(v)=\xi_1 v$ and $c_2(v)=\xi_2 v$. The values of Airy function $(Ai(x),~Bi(x))$ and its derivatives $(Ai'(x),~Bi' (x))$ at $x=0$ are given by, $Ai(0)=\frac{1}{3^{2/3}\Gamma(2/3)}$, $Bi(0)=\frac{1}{3^{1/6}\Gamma(2/3)}$, $Ai'(0)=-\frac{1}{3^{1/3}\Gamma(1/3)}$ and $Bi'(0)=\frac{3^{1/6}}{\Gamma(1/3)}$. Here, $\Gamma$ denotes the usual gamma function. The solutions of the above quadratic Eq. (\ref{44}) can be written as:
\ben
Y_{0{\pm}}=\frac{1}{8~3^{1/3}(\xi_1+\sqrt{3}\xi_2)}\Bigg[3\Gamma(\frac{2}{3})(2+\phi_v^2)\pm\sqrt{\Gamma(-\frac{1}{3})\Big(64~3^{1/3}(\xi_1+\sqrt{3}\xi_2)+\Gamma(-\frac{1}{3})(2+\phi_v^2)^{2}\Big)}\Bigg].
\label{45}
\een
Here we denote $Y_{0+}$ for the positive root and $Y_{0-}$ for the negative root. As we have mentioned earlier we are interested in real solutions only, for this, the condition is
\ben
\Gamma(-\frac{1}{3})\Big(64~3^{1/3}(\xi_1+\sqrt{3}\xi_2)+\Gamma(-\frac{1}{3})(2+\phi_v^2)^{2}\Big)\geq 0
\label{46}
\een
must hold true.

Now to achieve a locally naked singularity we must have one of the solutions to be positive but for globally naked singularity we should have both the solutions to be positive, i.e.,  $Y_{0\pm}> 0$. The condition in terms of $\phi_v^2$ can be written as
\ben
(2+\phi_v^2)^2\geq-64~\frac{3^{1/3}}{\Gamma(-\frac{1}{3})}\Big(\xi_1+\sqrt{3}\xi_2\Big).
\label{47}
\een

\begin{figure*}
\begin{minipage}[b]{0.4\linewidth}
\centering
 \begin{subfigure}[b]{0.9\textwidth}
    \includegraphics[width=7cm]{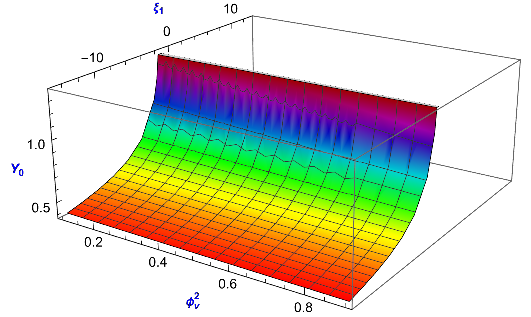}
        \caption{ $\xi_2=1$}
        \label{1a}
    \end{subfigure}
\end{minipage}
\hspace{2cm}
\begin{minipage}[b]{0.4\linewidth}
\centering
 \begin{subfigure}[b]{0.9\textwidth}
    \includegraphics[width=7cm]{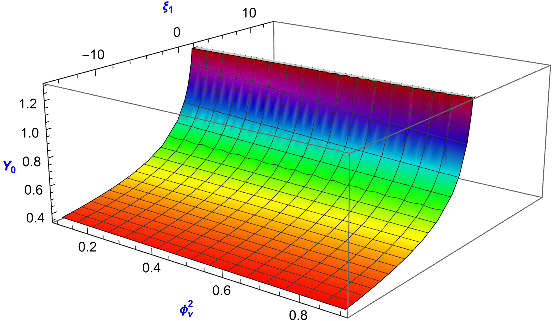}
        \caption{ $\xi_2=-1$}
        \label{1b}
    \end{subfigure}
\end{minipage}
\caption{Variation of $Y_0$ with $\phi_v^2$ for $\xi_2=$ constant and $\xi_1$ variable}
\end{figure*}

\begin{figure*}
\begin{minipage}[b]{0.4\linewidth}
\centering
 \begin{subfigure}[b]{0.9\textwidth}
    \includegraphics[width=7cm]{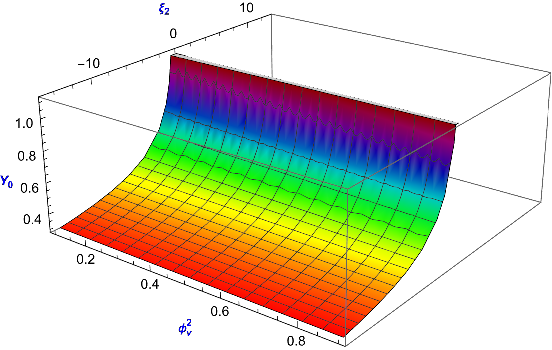}
        \caption{$\xi_1=1$}
        \label{2a}
    \end{subfigure}
\end{minipage}
\hspace{2cm}
\begin{minipage}[b]{0.4\linewidth}
\centering
 \begin{subfigure}[b]{0.9\textwidth}
    \includegraphics[width=7cm]{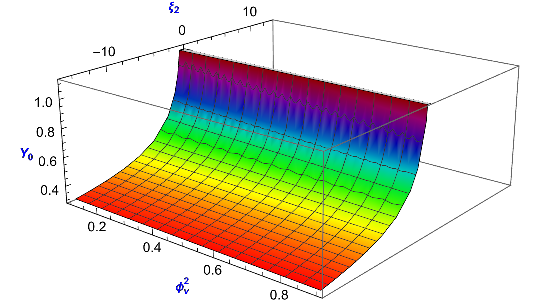}
        \caption{$\xi_1=-1$}
        \label{2b}
    \end{subfigure}
\end{minipage}
\caption{Variation of $Y_0$ with $\phi_v^2$ for $\xi_1=$ constant and $\xi_2$ variable}
\end{figure*}

Let us now check the occurrence of a naked singularity through a graphical analysis of Eq. (\ref{45}). As we know the value of $\phi_v^2$ should lie in between $0~\text{and}~1$ we vary it accordingly. The arbitrary constants $\xi_1~\text{and}~\xi_2$ can have both positive and negative values. So, we plot two graphs. First, we consider $\xi_2=1$, i.e., a positive value and then vary $\xi_1$ from negative to positive value [Fig. (\ref{1a})]. Second, we chose $\xi_2=-1$, i.e., a negative value and again varied $\xi_1$ from positive to negative value [Fig. (\ref{1b})]. Next, we interchange the position of $\xi_1~\text{and}~\xi_2$ and plot two more graphs [Figs. (\ref{2a}) and (\ref{2b})]. Based on the aforementioned graphs, it is evident that the value of $Y_0$ consistently exhibits positivity, hence indicating the presence of a {\it global naked singularity}. In Case-1, the presence of a naked singularity is seen instead of a black hole, which is mostly explained by the gravitational standpoint rather than being influenced by K-essence as a dark energy theory.\\

{\bf Case 2:}
 It has been noted that, in the limiting case, i.e., at $v\rightarrow 0$ and $r\rightarrow 0$ the value of $Y_{0\pm}$ becomes exactly the same as in case-1. We would get similar conditions (\ref{44}) at the central singularity. The condition for the real solution of $Y_{0}$ becomes the same as mentioned in (\ref{47}). Clearly, this is because of the choice of the constants $\b_1$ and $\b_2$. As a consequence, in this case, also we get a global naked singularity rather than a black hole.\\
 
{\bf Case-3:}
Following the similar approach to Case 1, we compute the value of $Y(=v/r)$ at $r\rightarrow 0, v\rightarrow 0$. In this case, we get

\ben
\frac{2}{Y_0}=\lim \limits_{\substack{%
    v \to 0\\
    r \to 0}}\Bigg(1-\frac{2g_1\Big[-1+2\b_2(1+\sqrt{1-\phi_v^2})\Big]+g_2\Big[-1+2\b_2(1-2n+\sqrt{1-\phi_v^2})\Big]}{2rg_2\b_2 n(\omega-1)\Big[1+\b_2\Big(n-2(1+\sqrt{1-\phi_v^2})\Big)\Big]}\Bigg).~~~~~~
\label{48}
\een

The limit of the expression $\frac{2}{Y_0}$ diverges as $r$ approaches zero. To obtain an accurate solution for the aforementioned equation, it is necessary for the mass function to explicitly depend on both $v$ and $r$. However, in this particular scenario, this condition is not satisfied. Thus far, we have not attained any instances of a collapsing situation. However, variation of a mass parameter may be seen. Based on the variation of the mass parameter with $\phi_{v}^{2}$ and time ($v$), we may conclude that it is possible to encounter either a bouncing or accelerating scenario, as stated below.\\

{\bf Case-4:}
Similar to Case-3 here we can not study the gravitational collapse as the mass parameter (\ref{41}) does not depend on $r$. But the variation of a mass parameter can be observed here too.

\begin{figure*}
\begin{minipage}[b]{0.4\linewidth}
\centering
 \begin{subfigure}[b]{0.9\textwidth}
    \includegraphics[width=6.5cm]{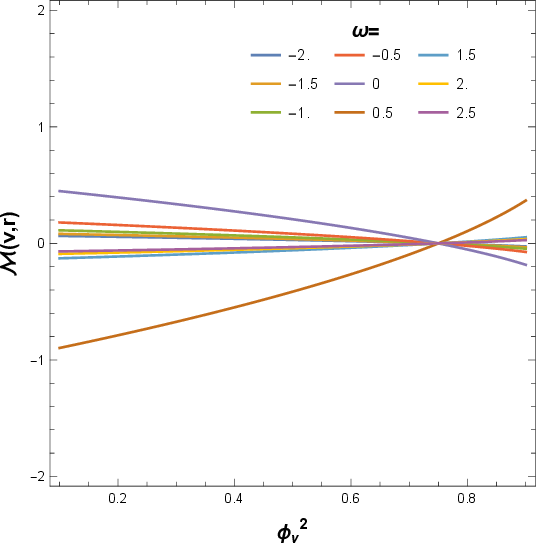}
        \caption{$\M(v,r)$ vs $\phi_v^2$}
        \label{4a}
    \end{subfigure}
\end{minipage}
\hspace{2cm}
\begin{minipage}[b]{0.4\linewidth}
\centering
 \begin{subfigure}[b]{0.9\textwidth}
    \includegraphics[width=6.5cm]{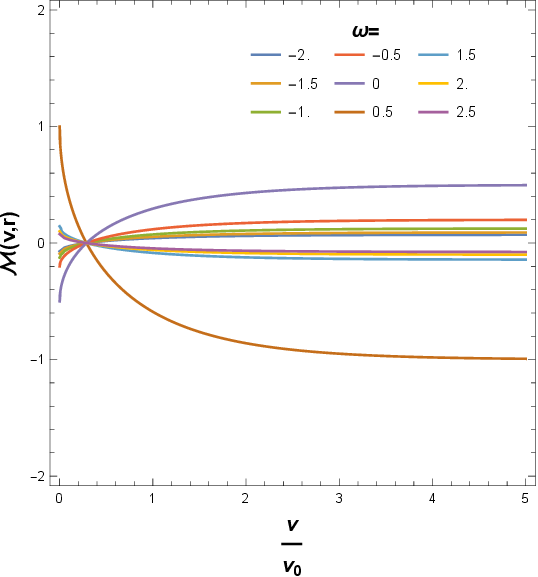}
        \caption{$\M(v,r)$ vs $v/v_0$}
        \label{4b}
    \end{subfigure}
\end{minipage}
\caption{Variation of mass parameter (Case 4: $\M(v,r)$) for different EoS parameter ($\omega$)}
\end{figure*}

Now, using graphical analysis we shall discuss the variations in the mass parameters Eqs. (\ref{38}) and  (\ref{41}) for case-3 and case-4. Basically, case-3 and case-4 present fascinating scenarios that can be investigated as follows. There is a clear dependence of $\phi_v^2$ on the mass parameter $\M(v,r)$ in (\ref{38}) and (\ref{41}). When the mass parameter is plotted against $\phi_v^2$ for various values of $\omega$, the resulting graphs are denoted as Figs. (\ref{3a}) and (\ref{4a}) for instances case-3 and case-4 respectively. We can observe that for every EoS parameter ($\omega$), the mass function reduces as $\phi_{v}^{2}$  increases, and at a particular value of $\phi_{v}^{2}~(=0.75)$, the mass function coincides with zero in two cases and varies again. It is important to note that the values of the kinetic energy of the K-essence scalar field ($\phi_{v}^{2}$) are $0<\phi_{v}^{2}<1$. The above feature can be explained as follows:

\begin{figure*}
\begin{minipage}[b]{0.4\linewidth}
\centering
 \begin{subfigure}[b]{0.9\textwidth}
    \includegraphics[width=6.5cm]{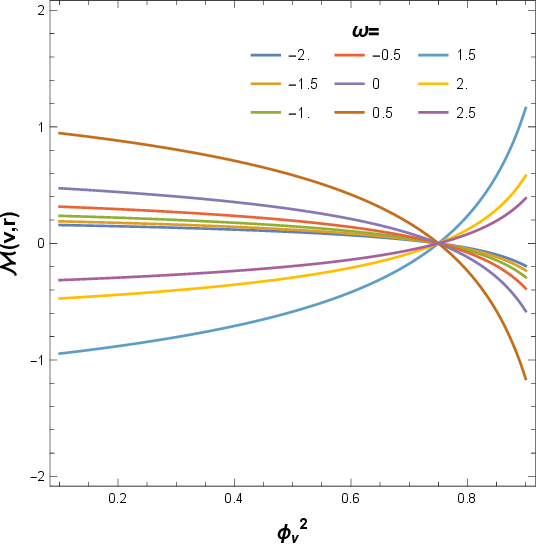}
        \caption{$\M(v,r)$ vs $\phi_v^2$}
        \label{3a}
    \end{subfigure}
\end{minipage}
\hspace{2cm}
\begin{minipage}[b]{0.4\linewidth}
\centering
 \begin{subfigure}[b]{0.9\textwidth}
    \includegraphics[width=6.5cm]{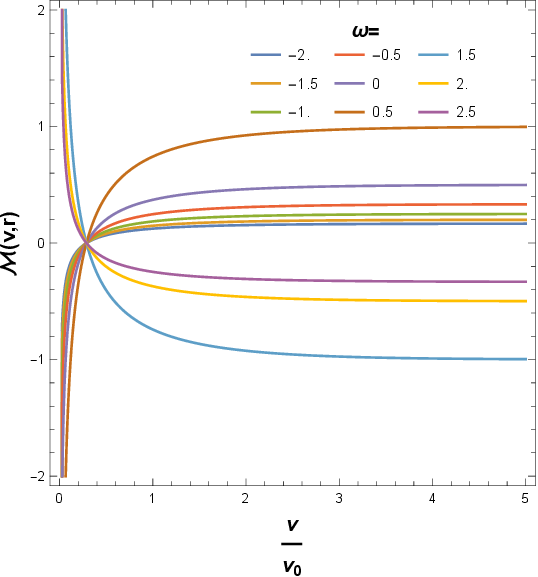}
        \caption{$\M(v,r)$ vs $v/v_0$}
        \label{3b}
    \end{subfigure}
\end{minipage}
\caption{Variation of mass parameter (Case 3: $\M(v,r)$) for different EoS parameter ($\omega$)}
\end{figure*}

The K-essence idea is also used in the study of the impact of dark energy throughout the cosmos~\cite{Scherrer,Vikman,Yoo}. Furthermore, it has been shown by several observations, such as those conducted by WMAP, Planck collaborations, and IA Supernovae, etc.~\cite{Riess,Perlmutter,Komatsu,Planck1,Planck2,Planck3}, that the estimated present value of dark energy density is around $0.74$. The kinetic energy of the K-essence scalar field ($\phi_{v}^{2}$) can be employed as a representation of dark energy density in the unit of critical density~\cite{gm1,gm2,gm3,gm4,gm8}. 

So, at this point ($\phi_{v}^{2}=0.75$), the mass has been completely transformed into energy, which has caused spacetime to change into Minkowski spacetime. As a result, it is possible to argue that the universe is dominated by dark energy and that spacetime is of the Minkowski type at the point $0.75$. This phenomenon occurs exclusively when observing the cosmos from an asymptotically flat perspective. It is evident that we are now studying the emergent Vaidya spacetime as our fundamental metric, which possesses a dynamical horizon~\cite{Hayward,Ashtekar1,Ashtekar2,Ashtekar3}, detail explained by Manna et al. \cite{gm5,gm6,gm10}. This dynamical horizon comes through the mass function $\M(v,r)$ affected by the kinetic energy of the K-essence scalar field ($\phi_{v}^{2}$). The existence of dark energy implies that the expansion of the universe is accelerating, as opposed to undergoing a collapse.

Once again, there is a constraint on the value of $\phi_v^2$ where $0<\phi_{v}^{2}<1$. The variable $v$ may be represented as $\phi_v^2=e^{-v/v_0}\theta(v)$, as stated in the Ref. \cite{gm5}. When the mass parameter is plotted against the ratio of time ($v$) to the reference time ($v_0$), we find Figs. (\ref{3b})  and  (\ref{4b}) for cases-3 and case-4, respectively. The constant $v_0$ is chosen to parameterize the unit of $v$. Additionally, it is stated that Figs. (\ref{4a}) and (\ref{4b}) illustrate the variations in $\M(v,r)$ as a function of $\phi_v^2$ and $v/v_0$ respectively, assuming that $g_1,~g_2,~n$ are set to a value of unity for the sake of simplicity. The variation of the mass parameters with respect to time ($v$) in units of $v_0$ may be observed from Figs. (\ref{3b}) and (\ref{4b}). Based on the analysis of Figs. (\ref{3b}) and (\ref{4b}), it is observed that the mass parameter undergoes a growth, either positive or negative, depending on the values of the EoS parameter $\o$, subsequent to reaching the value $v/v_0 = 0.287682$.

Based on the aforementioned graphs, specifically Figs. (\ref{3a}), (\ref{4a}), (\ref{3b}) and (\ref{4b}), we generate two tables, denoted as Tables-\ref{Table1} and \ref{Table2}. Table-\ref{Table1} presents an overview of the variations in the mass parameters $\M(v,r)$ under different values of $\o$ for case-3 and case-4, specifically at $\phi_v^2=0.5$ and $\phi_v^2=0.9$. Table-\ref{Table2} presents the mass parameters $\M(v,r)$ for various values of $\o$ in case-3 and case-4, specifically when $\frac{v}{v_0}=2$.

\begin{table}
\begin{center}
\resizebox{8cm}{!}{
\begin{tabular}{ |c|c|c|c|c|c|} 
\hline
\multirow{3}{4em}{~~~~{\bf $\o$}} &\multirow{3}{4em}{~~{\bf Epoch}} &  \multicolumn{4}{c}{{\bf $\M(v,r)$}} \\
\hline
 & &  \multicolumn{2}{c|}{$\phi_v^2=0.5$} &   \multicolumn{2}{c}{$\phi_v^2=0.9$}  \\
 \hline
 & & Case 3 & Case 4 & Case 3  & Case 4\\
\hline
\hline
 -2 & Dark Energy era & 0.0976 & 0.0296 & -0.1937 & -0.0262 \\
 \hline
-1.5 & Dark Energy era & 0.1171 & 0.0376 & -0.2324 & -0.0334 \\
 \hline
 -1.0 & Dark Energy era & 0.1464 & 0.0518 & -0.2905 & -0.0459 \\
 \hline
 -0.5 & Dark Energy era & 0.1953 & 0.0828 & -0.3874 & -0.0735 \\
 \hline
 0 & Dust era & 0.2929 & 0.2071 & -0.5811 & -0.1838 \\
 \hline
0.5 & Early Universe & 0.5858 & -0.4142 & -1.16223 & 0.3675 \\
 \hline
 1.0 & Stiff Fluid & Diverging & -0.1035 & Diverging & 0.0919 \\ 
 \hline
 1.5 & & -0.5858 & -0.0592 & 1.1623 & 0.0525 \\
 \hline
\end{tabular}}
\end{center}
\caption{Table of $\M(v,r)$ for different $\o$ for Cases 3 and 4 at $\phi_v^2=0.5$ and $\phi_v^2=0.9$.}
\label{Table1}
 \end{table}

\begin{table}[h]
\begin{center}
\resizebox{8cm}{!}{
\begin{tabular}{ |c|c|c|c|c|c|} 
\hline
\multirow{3}{4em}{~~~~{\bf $\o$}} &\multirow{3}{4em}{~~{\bf Epoch}} &  \multicolumn{2}{c}{{\bf $\M(v,r)$}} \\
\hline
 & &  \multicolumn{2}{c|}{$\frac{v}{v_0}=2$} \\
 \hline
 & & Case 3 & Case 4 \\
\hline
\hline
 -2 & Dark Energy era & 0.1541 & 0.0614  \\
 \hline
-1.5 & Dark Energy era & 0.1849 & 0.0781  \\
 \hline
 -1.0 & Dark Energy era & 0.2311 & 0.1075 \\
 \hline
 -0.5 & Dark Energy era & 0.3082 & 0.1719\\
 \hline
 0 & Dust era & 0.4622 & 0.4299\\
 \hline
0.5 & Early Universe & 0.9246 & -0.8597 \\
 \hline
 1.0 & Stiff Fluid & Diverging & -0.2149  \\ 
 \hline
 1.5 & & -0.9246 & -0.1228 \\
 \hline
\end{tabular}}
\end{center}
\caption{Table of $\M(v,r)$ for different $\o$ for Cases 3 and 4 at $\frac{v}{v_0}=2$.}
\label{Table2}
 \end{table}

According to the data shown in Table \ref{Table1}, it is evident that all the mass parameters exhibit positive values until the dust period, specifically at $\phi_{v}^{2}=0.5$. After that for case-4, it is seen all values of the mass parameter are negative and for case-3 at $\o=0.5$, the mass parameter is positive, at $\o=1.0$, it is diverging, and at $\o=1.5$, it is again negative. On the other hand, when $\phi_{v}^{2}=0.9$, the majority of the mass parameter values exhibit negativity in both scenarios, but a subset of values demonstrate positivity. In both scenarios, namely before and subsequent to the current value of the dark energy density ($\phi_{v}^{2}~(=0.74)$), some mass parameters exhibit negativity. The occurrence of unexpected properties in our scenarios can be ascribed to the presence of negative masses, which may indicate the probable existence of a gravitational dipole~\cite{Bondi,Miller} within our universe, with signs of future expansion. In this discourse, we explore the concepts of positive and negative masses in contemporary context.

Though we acknowledge the existence of positive physical masses \cite{Schoen1,Schoen2,Witten}, the idea of negative mass has been extensively studied in several scientific articles ~\cite{Miller,Vertogradov,Bondi,Bonnor,Farnes} across different settings. It is pointed out that Bondi and Bonnor \cite{Bondi,Bonnor} in their articles elaborately discussed the existence of negative mass along with the positive mass. In his work, Miller~\cite{Miller} provides a description of negative mass-lagging cores within the context of the Big Bang. The author stated that for very high values of the geometrically specified coordinate $r$, the mass is positively valued, while, at considerably small values of $r$, the mass is negatively valued. The area of spacetime characterized by negative mass has local features that resemble with those observed in the negative-mass Vaidya solution. One may endeavour to establish a non-singular past for the given hypersurface by considering a possible scenario in which we have a celestial object composed exclusively of particles that undergo decay, resulting in the emission of photons that subsequently dissipate. The description of the spacetime outside the contracting star is effectively explained by a Vaidya solution, whereby the relationship between the mass and the outgoing null coordinate $v$ is established based on the rate of particle disintegration. 

The negative energy dark matter paradigm suggests dark matter may be composed of negative energy, rather than the standard notion of unseen matter \cite{Siagian}. Negative matter has also been treated as Unified Dark Matter-Energy \cite{Chang}. Therefore, negative mass cosmology may be a pioneer in the study of Dark matter and Dark Energy.

Bondi~\cite{Bondi} has identified a notable phenomenon known as the ``gravitational dipole'', which is defined by the presence of both a positive mass and a negative mass. Positive mass exhibits attractive behaviour towards negative mass, whereas negative mass demonstrates repulsive behaviour towards positive mass. Consequently, both masses experience acceleration in a corresponding direction, resulting in the negative mass pursuing the positive mass, leading to the attainment of exceptionally high velocities by the dipole. Also, it should be noted that in our generalized emergent Vaidya spacetime (\ref{19}), the removal of the $\phi_{v}^{2}$ term and the substitution $\M(v,r)\equiv m(v)$ as well as not considering $f(\R,\T)$ gravity, which corresponds to the conventional Vaidya spacetime, lead us back to the investigation conducted by Miller in his published work~\cite{Miller}, which established the existence of both the positive and negative masses.\\

Table \ref{Table2} also presents the positive and negative aspects of the mass parameter, contingent upon the values of the EoS parameter. This observation further suggests the presence of dark energy and gravitational dipole across the many epochs of the cosmos, as shown by the presence of both negative and positive mass parameters, which are contingent upon the specific value of the EoS parameter. It should be also noted that a negative value of the EoS parameter indicates a world that is predominantly dominated by dark energy. Typically, the EoS parameter value for a dark energy-dominated universe is considered to be -1 and a larger negative value signifies a greater dominance of dark energy in the universe. The EoS parameter zero is often known as the dust or matter-dominated epoch. A value of $0.5$ indicates scenarios pertaining to the early universe, whereas a value of $1.0$ corresponds to the stiff fluid era. The presence of a negative mass parameter, coupled with a positive EoS parameter, suggests the possibility of the occurrence of negative mass during the early stages of the cosmos. On the other hand, the appearance of a gravitational dipole and dark energy characterized by negative EoS parameters demonstrate the phenomenon of late-time acceleration in the cosmos. This implies that in our model, both negative and positive masses simultaneously exist, given the specific choices of $f(\R, \T)$ as outlined in Eqs. (\ref{36}) and (\ref{39}), subject to certain fixed constants. So, we can say that under the concept of a gravitational dipole, there exists a dark energy-dominated universe, implying that the world is accelerating.

These phenomena (i.e., the presence of positive and negative masses) may be further elucidated via the lens of negative-mass cosmology, which offers a unified explanation for dark matter and dark energy~\cite{Farnes}. This work suggest that the existence of negative and positive masses in cosmology is indicative simultaneous presence of dark matter and dark energy. It is to be noted that we did not use the FLRW metric as a foundation in our work. Instead, we employed the K-essence generalized Vaidya metric for our investigations. Both the positive and negative masses in different epochs of the cosmos have been found for Cases 3 and 4, which points to the coexistence of dark matter and dark energy. It has been argued by Scherrer~\cite{Scherrer} that the purely kinetic K-essence may be utilized as a unified explanation of dark matter and dark energy, and we encompass it in our modified gravity framework (\ref{17}). Based on the findings FROM our investigation, it is possible to conclude that the presence of both the negative and positive masses is indicative of the presence of dark matter and dark energy in a united form.

\section{Strength of Singularity}
To measure the strength of singularity we remember the works of \cite{Tipler,gm6}. The condition for strong singularity $(r=v=\l=0)$ in our case is given by
\ben
\mathcal{A}=\lim\limits_{\l\to 0}\l^{2}\psi=\lim\limits_{\l\to 0}\l^{2}\bar{R}_{\mu\nu}K^{\mu}K^{\nu}>0,
\label{49}
\een
where $\bar{R}_{\mu\nu}$ is Ricci tensor and we define $\psi=\bar{R}_{\mu\nu}K^{\mu}K^{\nu}$ as a scalar of the K-essence emergent Vaidya spacetime and $\lambda$ is the affine parameter. Mention the fact that this particular scalar, $\psi$, is not a K-essence scalar field and is also not linked with the background gravitational metric $g_{\mu\nu}$. With the help of \cite{Mkenyeleye} we can show that 
\ben
\mathcal{A}=\lim \limits_{\substack{%
    \lambda \to 0}}\lambda^2 \psi=\frac{1}{4} Y_0^2 (2\dot{\M}_0(v,r))
\label{50}
\een
where $\M_0(v,r)=\lim \limits_{\substack{%
    v \to 0\\
    r \to 0}}\M(v,r)$ and 
$\dot{\M_0}(v,r)=\lim \limits_{\substack{%
    v \to 0\\
    r \to 0}}\frac{\partial \M(v,r)}{\partial v}$.

It should be emphasized that the criteria mentioned above can only be used to quantify the strength of singularity in the first two circumstances. Because the other two examples, cases 3 and 4, are unable to provide the collapse scenario. {The cases} 1 and 2 provide the same results in the limiting condition. As a result, we simply analyze the strength of singularity in case 1. Using the value of $\M(v,r)$ from Eq. (\ref{30}) we can write 
\ben
\mathcal{A}=\frac{Y_0^2}{2\Gamma(2/3)}\Big(\frac{\xi_1}{3^{2/3}}+\frac{\xi_2}{3^{1/6}}\Big).
\label{51}
\een

Since $Y_0^2>0$, it is clear that the signature of the above equation does not depend on the collapsing parameter $Y_0$, rather it depends on the previously defined parameters $\xi_1$ and $\xi_2$. Therefore, the condition for which the singularity will be strong is given by
\ben
\frac{1}{2\Gamma(2/3)}\Big(\frac{\xi_1}{3^{2/3}}+\frac{\xi_2}{3^{1/6}}\Big)>0
\label{52}
\een
and that for weak singularity is 
\ben
\frac{1}{2\Gamma(2/3)}\Big(\frac{\xi_1}{3^{2/3}}+\frac{\xi_2}{3^{1/6}}\Big)\leq 0.
\label{53}
\een

\section{Discussion with Conclusion}

In this section, we provide an overview of our research findings, which may be summarised as follows: The emergent $f(\R,\T)$ gravity in the K-essence geometry was initially taken into consideration. The Lagrangian employed in this study is a non-canonical Lagrangian of the DBI type, as shown by Eq. (\ref{8}). Additionally, the modified field equation associated with this theory, represented by Eq. (\ref{17}), corresponds to the emergent energy-momentum tensor ($\T_{\mu\nu}$) as indicated in Eq. (\ref{14}). This part has been summarised in section II. On the other hand, we have also taken into account the emergent Vaidya spacetime as our primary metric (\ref{19}), while specifying the mass function in (\ref{20}) and the energy-momentum tensor in (\ref{22}). In accordance with the reference~\cite{gm6}, several matter-dependent parameters, namely $\bar{\gamma}$, $\bar{\rho}$, $\bar{p}$, have been introduced. These parameters correspond to the energy density associated with the emergent Vaidya null radiation, as well as the energy density and pressure of the emergent perfect fluid, respectively. Note that the mass parameter in this context has been considered as a function of $r$ and $v$, where $v$ represents the null coordinate associated with the Eddington advanced time. The value of $r$ decreases towards the future along a ray that is associated with a constant value of $v$. The temporal variation of the emergent mass function ($\M(v,r)$) arises from the generalized Vaidya mass ($m(v,r)$) in the background, as well as the kinetic energy associated with the K-essence scalar field ($\phi_{v}^{2}$). The functional form of $f(\R,\T)$ is seen as an additive composition of two functions, namely $f_1(\R)$ which only relies on the Ricci scalar ($\R$) and $f_2(\T)$ which solely relies on the trace of the energy-momentum tensor ($\T$). There may be alternative forms of $f(\R,\T)$ that might be considered in future discussions. By referencing the field equations in (\ref{24}), (\ref{25}) and (\ref{26}), we go to the subsequent stage of our study. In our analysis, we focus on the $(22)$ or $(33)$ component of the field equation in order to create a comprehensive model. 

In the subsequent section, section III, we analyze four models of the $f(\R,\T)$ theory and determine the solution for the generalized mass parameter, denoted as $\M(v,r)$, for each of the above cases: Case 1 (Eq. \ref{30}), Case 2 (Eq. \ref{35}), Case 3 (Eq. \ref{38}) and Case 4 (Eq. \ref{41}). The solutions for Case 1 and Case 2 give identical findings given the specified conditions on $\b_1$, $\b_2$, $g_1$, and $g_2$. Whereas the solutions in case-3 and case-4 exhibit similarities under certain particular circumstances on the parameters. In our proposed model, the mass functions exhibit an additional K-essence-related component, denoted as $(\phi_v^2)$, which distinguishes it from the general case. This unique characteristic encourages us to investigate the remarkable behaviour of the mass function, as discussed in Section IV.

In Section IV we study the underlying physics of the mass parameters obtained in the previous section. The mass functions in Case-1 and Case-2 have an explicit dependency on $r$, allowing us to study the gravitational collapse and the fate of singularity. The limiting conditions on the tangent of the radial null geodesic (\ref{43}) tell us that the singularity formed in our case is naked for any value of $\phi_v^2$. Figs. (\ref{1a}),  (\ref{1b}), (\ref{2a}) and (\ref{2b}) support the statement. In this procedure, we check all the possible values of the constants $\xi_1$ and $\xi_2$. On the way, we have established the condition (\ref{47}) for which the solution of Eq. (\ref{45}) is real. It is important to mention that Ray et al. \cite{gm7}  thoroughly discussed the possibility of a naked singularity by utilizing observable data and referencing several scholarly works. Also, it is important to mention that Nodehi et al.~\cite{Nodehi} have shown that the empirical limitations pertaining to the dimensions and circularity of the $M87^{\star}$ shadow do not exclude the potential existence of a naked singularity inside this compact object, which may also contain the gravitomagnetic monopole.

The study of the mass function pertaining to Cases 3 and 4 yields surprising results. Initially, it should be noted that the solutions of the mass function do not explicitly depend on the variables $r$ or $v$. Consequently, these solutions are unable to offer any information about the singularity. However, their dependence relies on the K-essence term $\phi_v^2$, which exhibits a temporal dependency on the variable $v$. In the aforementioned scenarios, the mass parameters exhibit a non-constant nature, since they are implicitly dependent on time due to the presence of the kinetic energy term ($\phi_v^2$) associated with the K-essence scalar field. When we plot the mass function with $\phi_v^2$ for different $\o$ we get that in each case the mass function reaches to zero at a particular value of $\phi_v^2$, which is $0.75$ [Figs. (\ref{3a}) and (\ref{4a})]. Now according to the observation, we know that the present value of dark energy is about $0.74$. That means the K-essence term $\phi_v^2$ can be considered as the dark energy component of the universe~\cite{gm1,gm2,gm3,gm4,gm8}. The mass function vanishes at the point $0.75$. This is a clear indication that the mass has been converted into energy at this point indicating that the spacetime we are now is Minkowski. We should remember the fact that we started with a generalized Vaidya spacetime in the context of K-essence geometry. Surprisingly it is seen that this phenomenon occurs within the context of flat spacetime in the current era of our universe, specifically for a certain value of $\phi_v^2$. Again the condition that makes the spacetime (\ref{19}) physical is that $\phi_v^2$ can have a value between $0$ and $1$. This enables us to consider it as an exponential function of $v/v_0$ ($\phi_v^2=e^{-v/v_0}\theta(v)$), which provides us with the scope to study the variation of mass parameter with respect to the normalized time coordinate $v/v_0$ [Figs. (\ref{3b}) and (\ref{4b})]. The zero point of mass function arises at around $0.287682$ in the $v/v_0$ axis intending the similar behaviour of the mass function as achieved with  $\phi_v^2$. In these figures, we see the mass function sometimes becomes positive and sometimes negative depending upon the value of $\o$ we choose. Two Table-(\ref{Table1}) and Table-(\ref{Table2}) have been arranged to study the behaviour of the mass function in these two cases namely, case-3 and case-4. In conclusion, it is important to reiterate that the presence of a positive and negative mass function in this study may potentially signify the existence of a gravitational dipole across various stages of the universe. Each curve on these graphs encompasses both positive and negative values of the mass function, either before or succeeding the zero value. The potential existence of this gravitational dipole might perhaps account for the phenomenon of early inflation and the subsequent accelerated expansion of the cosmos. Alternatively, we might infer that the presence of negative and positive masses suggests the coexistence of dark matter and dark energy within the unified framework of our theory. The occurrence of naked singularity has been observed in case-1 and case-2, hence enabling the measurement of singularity strength simply in these instances. The strong and weak conditions are referenced in Eqs. (\ref{52}) and (\ref{53}).

It is important to highlight that the K-essence theory has the capability to be fully utilised, allowing for the concurrent exploration of both dark energies \cite{Scherrer,Vikman,gm1,gm2,gm3,Yoo} and a purely gravitational viewpoint~\cite{gm5,gm6,gm9,gm10}  without taking into account dark energy. Our findings, which indicate the existence of dark energy, a Minkowski universe for a fixed point, the gravitational dipole and the simultaneous presence of dark matter and dark energy, {all} are consistent with the proposed theory. Additionally, our research provides evidence for the existence of a global naked singularity by examining gravitational collapse just from a gravitational perspective, without considering the dark energy density represented by $\phi_v^2$. Based on our findings, it can be asserted that the cosmic censorship hypothesis is no longer applicable within the context of our specific scenario. Therefore, it can be concluded that the K-essence theory possesses the potential to serve as both a dark energy framework and a simply gravitational theory simultaneously, facilitating the examination of diverse phenomena within the realm of cosmology.

\section*{Acknowledgement}
A.P. and G.M. acknowledge the DSTB, Government of West Bengal, India for financial support through Grant Nos. 856(Sanc.)/STBT-11012(26)/6/2021-ST SEC dated 3rd November 2023. S.R. is thankful to the Inter-University Centre for Astronomy and Astrophysics (IUCAA), Pune, India for providing a Visiting Associateship under which a part of this work was carried out and he also gratefully acknowledges the facilities under ICARD, Pune at CCASS, GLA University, Mathura. The research by M.K. was carried out at Southern Federal University with financial support from the Ministry of Science and Higher Education of the Russian Federation (State contract GZ0110/23-10-IF).\\

{\bf Conflicts of interest:} The authors declare no conflicts of interest.\\

{\bf Data availability:} There is no associated data with this article, and as such, no new data was generated or analyzed in support of this research.\\

{\bf Declaration of competing interest:}
The authors declare that they have no known competing financial interests or personal relationships that could have appeared to influence the work reported in this paper.\\

{\bf Declaration of generative AI in scientific writing:} The authors state that they do not support the use of AI tools to analyze and extract insights from data as part of the study process.

\end{document}